%% file: SIGIR2019.tex
\documentclass[sigconf]{acmart}
\usepackage{multirow}
\usepackage{array}
\def\BibTeX{{\rm B\kern-.05em{\sc i\kern-.025em b}\kern-.08emT\kern-.1667em\lower.7ex\hbox{E}\kern-.125emX}}
    
\copyrightyear{2019} 
\acmYear{2019} 
\setcopyright{acmlicensed}
\acmConference[SIGIR '19]{Proceedings of the 42nd International ACM SIGIR Conference on Research and Development in Information Retrieval}{July 21--25, 2019}{Paris, France}
\acmBooktitle{Proceedings of the 42nd International ACM SIGIR Conference on Research and Development in Information Retrieval (SIGIR '19), July 21--25, 2019, Paris, France}
\acmPrice{15.00}
\acmDOI{10.1145/3331184.3331326}
\acmISBN{978-1-4503-6172-9/19/07}

\begin{document}

%
\title{FAQ Retrieval using Query-Question Similarity and\\BERT-Based Query-Answer Relevance}
%


 \author{Wataru Sakata}
 \orcid{1234-5678-9012}
 \affiliation{%
   \institution{LINE Corporation}
   }
 \email{wataru.sakata@linecorp.com}

 \author{Tomohide Shibata}
  \authornote{currently at Yahoo Japan Corporation}
 \affiliation{%
   \institution{Kyoto University}
  \institution{CREST, JST}
  }
 \email{shibata@nlp.ist.i.kyoto-u.ac.jp}

 \author{Ribeka Tanaka}
 \affiliation{%
   \institution{Kyoto University}
  \institution{CREST, JST}
 }
 \email{tanaka@nlp.ist.i.kyoto-u.ac.jp}

 \author{Sadao Kurohashi}
 \affiliation{%
  \institution{Kyoto University}
  \institution{CREST, JST}
  \institution{NII, CRIS}
}
 \email{kuro@nlp.ist.i.kyoto-u.ac.jp}
%
\renewcommand{\shortauthors}{Sakata, et al.}

%
\begin{abstract}
Frequently Asked Question (FAQ) retrieval is an important task where
the objective is to retrieve an appropriate Question-Answer (QA) pair
from a database based on a user's query. We propose a FAQ retrieval
system that considers the similarity between a user's query and a
question as well as the relevance between the query and an
answer. Although a common approach to FAQ retrieval is to construct
labeled data for training, it takes annotation costs. Therefore, we
use a traditional unsupervised information retrieval system to
calculate the similarity between the query and question.  On the other
hand, the relevance between the query and answer can be learned by
using QA pairs in a FAQ database. The recently-proposed BERT model is
used for the relevance calculation. Since the number of QA pairs in
FAQ page is not enough to train a model, we cope with this issue by
leveraging FAQ sets that are similar to the one in question.  We
evaluate our approach on two datasets. The first one is localgovFAQ, a
dataset we construct in a Japanese administrative municipality domain.
The second is StackExchange dataset, which is the public dataset in
English.  We demonstrate that our proposed method outperforms baseline
methods on these datasets.
\end{abstract}

%
%


%

%
\maketitle


\vspace{-1.5mm}
\section{Introduction} \label{sec:intro}
\input{intro}

\vspace{-1.5mm}

\section{Proposed Method} \label{sec:method}
\input{method}
\vspace{-1.5mm}

\section{Experiments and Evaluation}
\input{experiments}
\vspace{-1.5mm}

\section{Conclusion}
This paper presented a method for using query-question similarity and BERT-based query-answer relevance in a FAQ retrieval task.
By collecting other similar FAQ sets, we could increase the size of available QA data.
BERT, which has been recently proposed, was applied to capture the relevance between queries and answers. This method realized the robust and high-performance retrieval. 
The experimental results demonstrated that our combined use of query-question similarity and query-answer relevance was effective.
    We are planning to make the code and constructed dataset localgovFAQ publicly available\footnote{http://nlp.ist.i.kyoto-u.ac.jp/EN/index.php?BERT-Based\_FAQ\_Retrieval}.

\vspace{-1.5mm}

\section*{Acknowledgment}
This work was partly supported by JST CREST Grant Number JPMJCR1301, Japan.
\vspace{-1mm}

\bibliographystyle{ACM-Reference-Format}
{{\bibliography{SIGIR2019}}}

\end{document}

%% file: intro.tex

There are often frequently asked question (FAQ) pages with various information on the web.
A FAQ retrieval system, which takes a user's query and returns relevant QA pairs, is useful for navigating these pages.

In FAQ retrieval tasks, 
it is standard to check similarities of user's query ($q$) to a FAQ's question ($Q$) or 
to a question-answer (QA) pair~\cite{DBLP:journals/eswa/KaranS18}.
Many FAQ retrieval models use the dataset with the relevance label between $q$ and a QA pair. 
However, it costs a lot to construct such labeled data.
To cope with this problem, we adopt an unsupervised method for calculating the similarity between a query and a question.

Another promising approach is to check the q-A relevance
trained by QA pairs, 
which shows the plausibility of the FAQ answer for the given $q$. 
Studies of community QA use a large number of QA pairs for learning the q-A relevance ~\cite{DBLP:journals/corr/TanXZ15,DBLP:conf/acl/WuwLZ18}. 
However, these methods do not apply to FAQ retrieval task, because the size of QA entries in FAQ is not enough to train a model generally.
We address this problem by collecting other similar FAQ sets to increase the size of available QA data. 

\begin{figure}[t]
  \includegraphics[width=0.98\columnwidth]{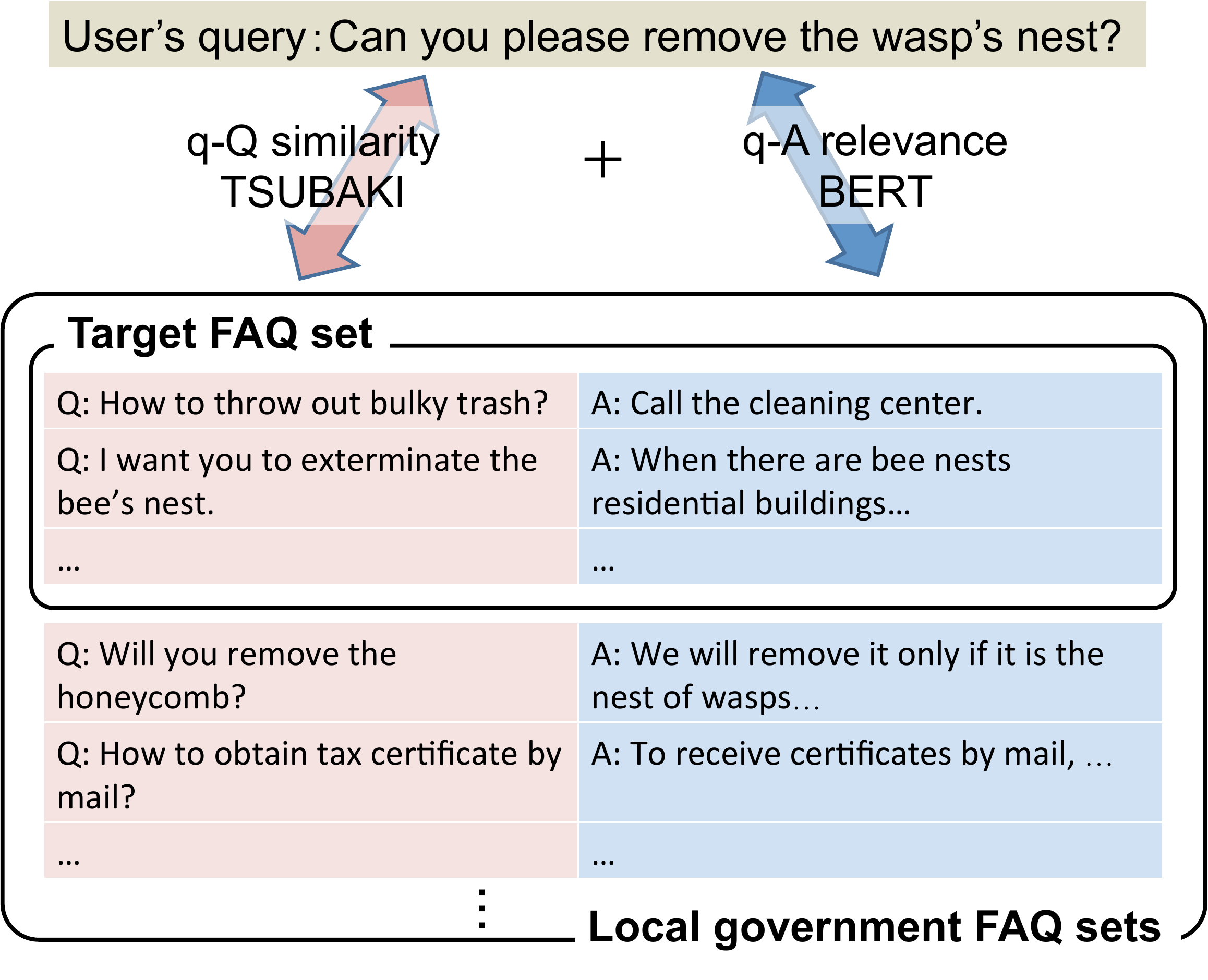}
   \vspace{-5mm}
  \caption{An overview of our proposed method.}
  \Description{An overview}
  \label{fig:model}
  \vspace{-4mm} 
\end{figure}

In this study, we propose a method that combines 
the q-Q similarity obtained by unsupervised model and the q-A relevance learned from the collected QA pairs.
Figure \ref{fig:model} shows the proposed model.
%
Previous studies show that neural methods (e.g., LSTM and CNN)
work effectively
in learning q-A relevance.
Here we use the recently-proposed model, BERT~\cite{devlin2018bert}.
BERT is a powerful model that applies to
a wide range of tasks and obtains the state-of-the-art results
on many tasks including GLUE \cite{W18-5446} and SQuAD \cite{D16-1264}.
An unsupervised retrieval system achieves high precision,
but it is difficult to deal with a gap between the expressions of $q$ and $Q$.
By contrast, since BERT validates the relevance between $q$ and $A$, 
it can retrieve an appropriate QA pair even if there is a lexical gap between $q$ and $Q$. 
By combining the characteristics of two models, we achieve a robust and high-performance retrieval system.

We conduct experiments on two datasets.
The first one is the localgovFAQ dataset,
which we construct to evaluate our model
in a setting where other similar FAQ sets are available.
It consists of QA pairs collected from Japanese local government FAQ pages
and an evaluation set constructed via crowdsourcing.
The second one is the StackExchange dataset~\cite{DBLP:journals/eswa/KaranS18},
which is the public dataset for FAQ retrieval tasks.
We evaluate our model on these datasets and show that the proposed method works effectively in FAQ retrieval.

%% file: method.tex
\subsection{Task Description}
\label{subsec:task}
We begin by formally defining the task of FAQ retrieval.
Here, we focus on local government FAQ as an example.
Suppose that the number of local government FAQ sets is $N$.
Our target FAQ set, $T_t$, is one of them.
When the number of QA entries in $T_t$ is $M$, $T_t$ is a collection of QA pairs $\{(Q_1, A_1), (Q_2, A_2), ...,(Q_M, A_M)\}$.
The task is then to find the appropriate QA pair $(Q_{i}, A_i)$
from $T_t$ based on a user's query $q$.
We use $T_1, T_2,...,T_N$ as our training data,
including the FAQ set $T_t$ of the target local government.


\vspace{-0.5mm}
\subsection{q-Q similarity by TSUBAKI}

We use TSUBAKI~\cite{Shinzato2008a} to 
compute q-Q similarity. TSUBAKI is an unsupervised retrieval engine based on OKAPI BM25~\cite{Okapi}. 
TSUBAKI accounts for a dependency structure of a sentence, not just its words, to provide accurate retrieval.
For flexible matching, it also uses synonyms automatically extracted from dictionaries
and Web corpus.
Here we regard $Q$ in each QA as a document and compute $Similarity(q, Q)$ for
the q-Q similarity.
%
%


\vspace{-0.5mm}
\subsection{q-A relevance by BERT}

We use BERT to compute q-A relevance.
BERT is based on the Transformer \cite{NIPS2017_7181} that effectively encodes an input text. It is designed to be pre-trained using a language model objective on 
a large raw corpus and fine-tuned for each specific task including sentence classification, sentence-pair classification, and question answering.
As it is pre-trained on a large corpus, BERT achieves high accuracy even if the data size of the specific task is not large enough. We apply BERT to a sentence-pair classifier for questions and answers. By applying the Transformer to the input question and answer, it captures the relevance between the pair.

The training data we use is
the collection of QA pairs from FAQ sets (see Sec.~\ref{subsec:task}).
For each positive example $(Q, A)$,
we randomly select $\bar{A}$
and produce negative training data $(Q, \bar{A})$.
On this data, we train BERT to solve the two-class classification problem:
$Relevance(Q, A)$ is 1 and $Relevance(Q, \bar{A})$ is 0,
where $Relevance(Q,A)$ stands for the relevance between $Q$ and $A$.

At the search stage, we compute $Relevance(q,A_i)$
$(i=1,\cdots, M)$ for the user's query $q$ and every QA pair in the target $T_t$.
QA pairs in a higher rank are used as search results.



\vspace{-0.5mm}
\subsection{Combining TSUBAKI and BERT}
\label{subsec:integrattion}

In order to realize robust and flexible matching, we combine the q-Q similarity by TSUBAKI and the q-A relevance by BERT.

When TSUBAKI's similarity score is high,
this is probably a positive case because the words in $q$ and $Q$ highly overlap with each other.
However, it is difficult to cope with the lexical gaps between  $q$ and $Q$.
On the other hand, since BERT validates the relevance between $q$ and $A$,
it can retrieve an appropriate QA pair even if there is a lexical gap between $q$ and $Q$.
To make use of these characteristics, we combine two methods as follows.
First, we take the ten-highest results of BERT's output.
For QA pairs whose TSUBAKI score gets a higher score than $\alpha$, we rank them in order of TSUBAKI's score.
For the others, we rank them in order of the score of $Similarity(q, Q)\times t + Relevance(q, A)$ where $t$ is a hyper-parameter.

TSUBAKI's score tends to be higher when the given query is longer. Hence, before taking the sum, we normalize TSUBAKI's score by using the numbers of content words and dependency relations in the query. We divide the original score by the following value.\footnote{
We do not normalize the BERT's score because it takes a value between 0 to 1.
}

\vspace{-0.5mm}
\begin{center}
  $Count(\mathit{ContentWords}) \times k_1 + Count(\mathit{DependencyRelations}) \times k_2$
\end{center}
\vspace{-1.0mm}

%% file: experiments.tex
We conducted our experiments on two datasets, \textit{localgovFAQ} and \textit{StackExchange}.
We constructed localgovFAQ dataset, as explained in Sec \ref{subsec:localgovfaq_construction}.
StackExchange dataset is constructed in the paper \cite{DBLP:journals/eswa/KaranS18}
by extracting QA pairs from the web apps domain of StackExchange and consists of 719 QA pairs. Each $Q$ has paraphrase queries, and the total number of queries is 1,250. All the models were evaluated using five-fold cross validation. In each validation, all the queries were split into training (60\%), development (20\%) and test (20\%). The task is to estimate an appropriate QA pair for each query $q$ among 719 QA pairs.
\vspace{-0.5mm}
\subsection{LocalgovFAQ Evaluation Set Construction}
\label{subsec:localgovfaq_construction}

\begin{figure}[t]
\begin{center}
\small
\begin{tabular}{p{7cm}} \hline
\vspace{-2.5mm}
\begin{itemize}
\setlength{\leftskip}{-0.5cm}
\item I'd like you to issue a copy of family register, but how much does it cost?
\item I'd like you to publish a maternal and child health handbook, but what is required for the procedure?
\item I'm thinking of purchasing a new housing, so I want to know about the reduction measure.
\item From which station does the pick-up bus of the Center Pool come out? \vspace{-1.0em}
\end{itemize}\\
\hline
\end{tabular}
\vspace{-3mm}
\caption{Examples of queries collected via crowdsourcing.} \label{tab:query}
\end{center}
\vspace{-4mm}
\end{figure}

Amagasaki-city, a relatively large city in Japan, was chosen as a target government, whose Web site has 1,786 QA pairs. First, queries to this government were collected using a crowdsourcing. Example queries are shown in Figure ~\ref{tab:query}. We collected 990 queries in total.

TSUBAKI and BERT output at most five relevant QA pairs for each query, and each QA pair was manually evaluated assigning one of the following four categories:

\begin{description}
\itemsep=0pt
\item[A] Contain correct information.
\item[B] Contain relevant information.
\item[C] The topic is same as a query, but do not contain relevant information.
\item[D] Contain only irrelevant information.
\end{description}

In general, information retrieval evaluation based on the pooling method has inherently a \textit{biased} problem.
To alleviate this problem, when there are no relevant QA pairs among the outputs by TSUBAKI and BERT, a correct QA pair was searched by using appropriate different keywords. If there are no relevant QA pair found, this query was excluded from our evaluation set. The resultant queries were 784. 
Since 20\% of queries were used for the development set, 627 queries were used for our evaluation.

\vspace{-0.5mm}
\subsection{Experimental Settings}
For the localgovFAQ dataset, MAP (Mean Average Precision), MRR (Mean Reciprocal Rank), P@5 (Precision at 5), SR@k (Success Rate)\footnote{Success Rate is
the fraction of questions for which at least one related question is
ranked among the top $k$. } and nDCG (normalized Discounted Cumulative
Gain) were used as our evaluation measures.
The categories A, B and C
were regarded as correct for MAP, MRR, P@5 and SR@k, and the evaluation level of
categories A, B and C was regarded as 3, 2 and 1 for nDCG, respectively. 
For the StackExchange dataset, MAP, MRR and P@5 were used, following Karan et al.  \cite{DBLP:journals/eswa/KaranS18} .

For Japanese, the pre-training of BERT was performed using Japanese Wikipedia, which
consists of approximately 18M sentences, and the fine-tuning was
performed using FAQs of 21 Japanese local governments, which consists of
approximately 20K QA pairs. The morphological analyzer Juman++\footnote{\url{http://nlp.ist.i.kyoto-u.ac.jp/EN/index.php?JUMAN++}} was
applied to input texts for word segmentation, and words were broken into subwords by applying BPE \cite{P16-1162}.
For English BERT pre-trained model, a publicly-available model was used\footnote{\url{https://storage.googleapis.com/bert_models/2018_10_18/uncased_L-12_H-768_A-12.zip}}. 
For the fine-tuning for StackExchange dataset, the training set $(q, Q, A)$ was divided into $(q, A)$ and $(Q, A)$.

In the localgovFAQ dataset, Bi-LSTM with attention \cite{DBLP:journals/corr/TanXZ15} was adopted as our baseline.
We also used model BERT$_\mathit{targetOnly}$, which is fine-tuned only with the target FAQ set, in order to test the effect of using other FAQ sets.
In the StackExchange dataset, CNN-rank in q-Q and q-QA settings, which is the neural FAQ retrieval model based on a convolutional neural network, was used, whose scores were from Karan et al. \cite{DBLP:journals/eswa/KaranS18}.
Furthermore, BERT (w/o query paraphrases) was adopted, where $(q, A)$ pairs were not used for BERT training, to see the performance
when no manually-assigned query paraphrases were available.
For both datasets, TSUBAKI was applied only in q-Q setting\footnote{We omit the TSUBAKI's results in the q-A setting and the q-QA setting as we got the worse scores than the q-Q setting.}.

For both BERT and Bi-LSTM models, 24 negative samples for one positive sample were used. For the coefficients explained in Sec. \ref{subsec:integrattion}, $k_1$ and $k_2$ were set to 4 and 2, respectively, and $\alpha$ and $t$ were set to 0.3 and 10 respectively using the development set.

\vspace{-0.5mm}
\subsection{Evaluation Results and Discussions}

Table \ref{tab:exp_result} shows an experimental result on localgovFAQ dataset. 
In q-A setting, BERT was better than the Bi-LSTM baseline, which indicates BERT was useful for this task.
Although the performances of TSUBAKI and BERT were almost the same in terms of SR@1, the performance of BERT was better than TSUBAKI in terms of SR@5, which indicates BERT could retrieve a variety of QA pairs.
The proposed method performed the best. This demonstrated the effectiveness of our proposed method.
The score of BERT was better than one of BERT$_\mathit{targetOnly}$, which indidates that using other FAQ sets is effective.

Table \ref{tab:exp_result_stackFAQ} shows an experimental result on StackExchange dataset. In the same way as the result on localgovFAQ, BERT performed well, and the proposed method performed the best in terms of all the measures. The performance of BERT was better than one of "BERT (w/o query paraphrases)", which indicates that the use of various augmented questions was effective.

\begin{figure}[t]
  \centering
\includegraphics[width=\linewidth]{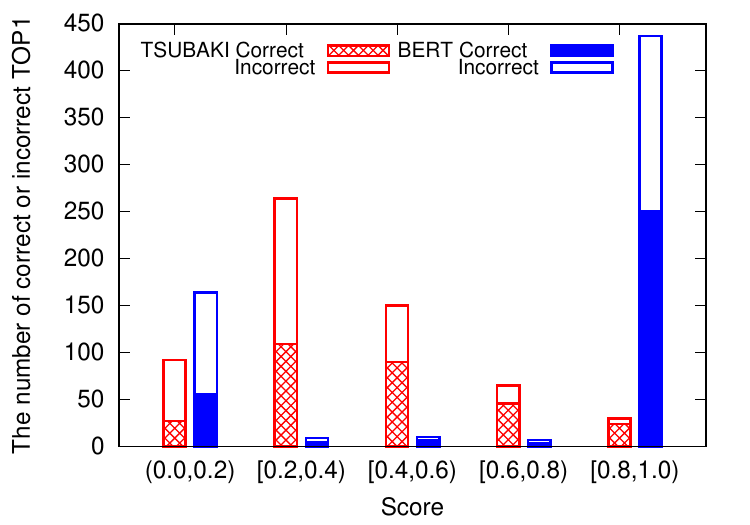}
\vspace{-6mm}
\caption{The number of queries to which the systems can or cannot output correct QA pair as TOP1 with respect to their scores.}
\vspace{-4mm}
  \Description{comparison}
  \label{fig:tb_score}
\end{figure}

Figure \ref{fig:tb_score} shows the performance of TSUBAKI and BERT on localgovFAQ according to their TOP1 scores.
From this figure, we can find that in the retrieved QA pair whose TSUBAKI score is high, its accuracy is very high. On the otherhand, there is a relatively loose correlation between the accuracy and BERT score. 
This indicates TSUBAKI and BERT have different characteristics, and our proposed combining method is reasonable.
\begin{table}[t]
\begin{center}
 \begin{tabular}{l|l|c@{\ }@{\ }c@{\ }@{\ }c@{\ }@{\ }c@{\ }c@{\ }c@{\ }c} \hline
   & \multicolumn{1}{c|}{Model} & MAP & MRR & P@5 & SR@1 & SR@5& NDCG\\\hline
   q-Q &  TSUBAKI &0.558  & 0.598 & 0.297 &0.504       &0.734   & 0.501\\\hline
   q-A &  Bi-LSTM &0.451  & 0.498 & 0.248 & 0.379       & 0.601 & 0.496\\
       &  BERT$_\mathit{targetOnly}$    & 0.559 & 0.610 & 0.285& 0.504       & 0.751 &0.526\\
       &  BERT    & 0.576 & 0.631 & 0.333& 0.509       & 0.810 &0.560\\\hline
   q-QA    &  Proposed& {\bf 0.647} & {\bf 0.705}& {\bf 0.357}& {\bf 0.612} & {\bf 0.841} & {\bf 0.621}\\\hline
\end{tabular}
  \caption{Evaluation result on the localgovFAQ dataset.} \label{tab:exp_result}
  \vspace{-7mm}
\end{center}
\end{table}

\begin{table}[t]
\begin{center}
 \begin{tabular}{l|l|c@{\ }@{\ }c@{\ }@{\ }c} \hline
   &\multicolumn{1}{c|}{Model} & MAP & MRR & P@5\\\hline
  q-Q &  CNN-rank & 0.79 & 0.77 & 0.63 \\
     & TSUBAKI & 0.698 & 0.669 & 0.638 \\\hline
  q-A  & BERT (w/o query paraphrases) & 0.631 & 0.805 & 0.546\\
   & BERT & 0.887 &  0.936 & 0.770\\\hline
  q-QA &  CNN-rank & 0.74 & 0.84 & 0.62 \\
     & Proposed & {\bf 0.897} & {\bf 0.942} & {\bf 0.776}\\\hline
\end{tabular}
\caption{Evaluation result on the StackExchange dataset.} \label{tab:exp_result_stackFAQ}
  \vspace{-7mm}
  \end{center}
\end{table}

\begin{table*}[t]
\small
\begin{center}
    \begin{tabular}{p{2.5cm}|@{}c@{}|p{4cm}|@{}c@{}|p{4cm}|@{}c@{}|p{4cm}} \hline
\multicolumn{1}{c|}{Query} & \multicolumn{2}{c|}{TSUBAKI} &
     \multicolumn{2}{c|}{BERT} & \multicolumn{2}{c}{Proposed method}\\\hline
    Is there a consultation desk for workplace harassment? &
\begin{tabular}[t]{@{}c@{}}
1\\$\times$
\end{tabular}
      &
      \begin{tabular}[t]{@{}p{4cm}@{}}
        Q: I'd like to have a career counseling.\\
        A: Consultation place: Amagasaki-city, ...
\end{tabular}&
\begin{tabular}[t]{@{}c@{}}
1\\\checkmark
\end{tabular}
      &
       \begin{tabular}[t]{@{}p{4cm}@{}}
         Q: I'd like to consult a lawyer for work-related problems. \\
         A: On specialized and sophisticated labor issues such as wages, dismissal, occupational accidents, ...
\end{tabular} &
\begin{tabular}[t]{@{}c@{}}
1\\\checkmark
\end{tabular}
      &
         \begin{tabular}[t]{@{}p{4cm}@{}}
           Q: Can we get a lawyer's labor counselor? \\
					A: On specialized and sophisticated labor issues such as wages, dismissal, occupational accidents, ...
\end{tabular}
    \\\hline

Where should I renew my license? &
\begin{tabular}[t]{@{}c@{}}
1\\$\times$
\end{tabular}
    &
      \begin{tabular}[t]{@{}p{4cm}@{}}
        Q: Where should I apply for medical staff licenses (new, corrected / rewritten, re-issued)?\\
        A: License application for doctors, dentists, public health nurses ...
      \end{tabular} &
\begin{tabular}[t]{@{}c@{}}
1\\$\times$
\end{tabular}
    &
      \begin{tabular}[t]{@{}p{4cm}@{}}
          Q: To update a bus ticket, do I have to go myself? 
			   \\A: In principle, please apply for the application by yourself. ...
      \end{tabular} &
\begin{tabular}[t]{@{}c@{}}
1\\\checkmark
\end{tabular}
    &
         \begin{tabular}[t]{@{}p{4cm}@{}}
  Q: Please tell me about the procedure of updating your driver's license.\\
  A: Regarding the renewal procedure of your driver's license ...
         \end{tabular}\\ \cline{2-7}

 &
\begin{tabular}[t]{@{}c@{}}
2\\\checkmark
\end{tabular}
    &
      \begin{tabular}[t]{@{}p{4cm}@{}}
        Q:  Please tell me about the procedure of updating your driver's license.\\
          A: Regarding the renewal procedure of your driver's license ...
      \end{tabular} &
\begin{tabular}[t]{@{}c@{}}
2\\$\times$
\end{tabular}
    &
      \begin{tabular}[t]{@{}p{4cm}@{}}
          Q: Can I file an agent application to renew my bus ticket?
        \\A: As a general rule, please apply for yourself. ...
      \end{tabular} &
\begin{tabular}[t]{@{}c@{}}
2\\$\times$
\end{tabular}
    &
         \begin{tabular}[t]{@{}p{4cm}@{}}
        Q: Can I file an agent application to renew my bus ticket?\\
          A:  As a general rule, please apply for yourself. ...
      \end{tabular}\\\hline

    Is there a place that we can use for practicing instruments?
&
\begin{tabular}[t]{@{}c@{}}
1\\$\times$
\end{tabular}
    &
      \begin{tabular}[t]{@{}p{4cm}@{}}
        Q: Where is the location of the polling place before the election's due date?
        \\A: There are three polling stations before the date in the city. ...
      \end{tabular} &
\begin{tabular}[t]{@{}c@{}}
1\\$\times$
\end{tabular}
    &
      \begin{tabular}[t]{@{}p{4cm}@{}}
        Q: Please tell me about Amagasaki City boys music corps.
			   \\A: "Amagasaki City Boys Music Club" includes a choir corps, a brass band, ...
      \end{tabular} &
\begin{tabular}[t]{@{}c@{}}
1\\$\times$
\end{tabular}
    &
         \begin{tabular}[t]{@{}p{4cm}@{}}
           Q: Please tell me about Amagasaki City boys music corps.\\
					A: "Amagasaki City Boys Music Club" includes a choir corps, a brass band, ...
      \end{tabular}\\\hline

\end{tabular}
\caption{Examples of system outputs and their manual evaluations. (\checkmark and $\times$ in the table mean correct and incorrect, respectively, where the evaluation categories A, B, and C are regarded as correct.)} \label{tab:examples}
\end{center}
\vspace{-9mm}
\end{table*}

Table \ref{tab:examples} shows the examples, translated from Japanese, of system outputs and correct QA pairs on localgovFAQ. In the first example, although TSUBAKI retrieved the wrong QA pair since there is a word "consultation" and "counseling" in the query and $Q$, BERT and the proposed method could retrieve a correct QA pair.
In the second example, the proposed method could retrieve a correct QA pair on the first rank although the first rank of TSUBAKI and BERT was wrong.

In the third example, no methods could retrieve a correct QA pair.
Although BERT could capture the relevance between a word "instruments" in the query and "music" in $A$, the retrieved QA pair was wrong.
The correct QA pair consists of $Q$ saying "Information on the facility of the youth center, hours of use, and closed day" and $A$ mentioning a music room as one of the available facilities in the youth center. To retrieve this correct QA pair, the deeper understanding of QA texts is necessary.

It takes about 2 seconds to retrieve QA pairs per query on localgovFAQ dataset by using 7 GPUs (TITAN X Pascal), and our model is practical enough.
For a larger FAQ set, one can use our method in a telescoping setting~\cite{Matveeva:2006:HAR:1148170.1148246}.